\documentstyle[aps,prb,eqsecnum]{revtex}
\begin{document}
\title{Dimer and N\'eel order-parameter fluctuations
in the spin-fluid phase \\
of the $s=1/2$ spin chain with first and second neighbor couplings}
\author{Yongmin Yu and Gerhard M\"uller}
\address{Department of Physics, The University of Rhode Island, Kingston,
Rhode Island 02881-0817}
\author{V.S. Viswanath}
\address{Solid State Division, Oak Ridge National Laboratory,
P.O. Box 2008, Oak Ridge, Tennessee 37831-6032}
\date{\today}
\maketitle

\begin{abstract}
The dynamical properties at $T=0$ of the one-dimensional (1D) $s=1/2$
nearest-neighbor ($nn$) $XXZ$ model with an additional isotropic
next-nearest-neighbor ($nnn$) coupling are investigated by means of the
recursion method in combination with techniques of continued-fraction
analysis.
The focus is on the dynamic structure factors $S_{zz}(q,\omega)$
and $S_{DD}(q,\omega)$, which describe (for $q=\pi$) the fluctuations of
the N\'eel and dimer order parameters, respectively.
We calculate (via weak-coupling continued-fraction analysis) the dependence
on the exchange constants of the infrared exponent, the renormalized
bandwidth of spinon excitations, and the spectral-weight distribution in
$S_{zz}(\pi,\omega)$ and $S_{DD}(\pi,\omega)$, all in the spin-fluid phase,
which is realized for planar $nn$ anisotropy and sufficiently weak $nnn$
coupling.
For some parameter values we find a discrete branch of excitations above the
spinon continuum. They contribute to $S_{zz}(q,\omega)$ but not to
$S_{DD}(q,\omega)$.
\end{abstract}
\pagebreak
\twocolumn

\section{Introduction}

Quantum many-body systems with competing interactions are apt to exhibit
ordering tendencies in the ground state of which no trace exists in the
presence of any one coupling alone and which are impossible to predict
on the basis of a classical model with both couplings.
Among the many different model systems where this phenomenon is manifest,
the Heisenberg antiferromagnet with
nearest-neighbor ($nn$) and next-nearest-neighbor ($nnn$) coupling on a
bipartite lattice is a prominent example.
For coupling strengths of comparable magnitude on the square lattice,
various ordering tendencies including N\'eel order, collinear order,
dimer order, twist order, and chiral order are in competition with each
other and with disordering tendencies such as embodied by the resonating
valence-bond state.\cite{R2D}

The one-dimensional (1D) version of this model had gained prominence many
years before, when the ground-state for one particular ratio of the coupling
constants was found to be a pure dimer state.\cite{MG69,B80}
Spontaneous dimerization is a true quantum phenomenon.
Several studies have succeeded in mapping out the zero-temperature phase
diagram of this model in one or the other extended parameter
space,\cite{SS81,H82,TH87,SGMK96} albeit with little emphasis on dynamical
properties.
In this paper we study the $T=0$ dynamics of the Hamiltonian
\begin{eqnarray}
H=\sum_{l=1}^N\{ J_{\perp}&&[S_l^xS_{l+1}^x+S_l^yS_{l+1}^y] \nonumber
\\ +&&J_zS_l^zS_{l+1}^z+J_2{\bf S}_l\cdot{\bf S}_{l+2}\} \;,
\label{I.1}
\end{eqnarray}
parametrized by the two coupling constants $\Delta =J_z/J_{\perp}$ and
$\Lambda=J_2/J_{\perp}$.
Extending the parameter space to cases with uniaxial anisotropy in the
$nn$ coupling offers the advantage that the model with $\Delta =\Lambda =0$,
which reduces to a system of free lattice fermions,\cite{LSM61,K62} can be
used as a convenient starting point for weak-coupling approaches.

A schematic representation of the $T=0$ phase diagram is
shown in Fig. \ref{1}. The free-fermion point ($\Delta=0,\Lambda=0$) is
located in the middle of the spin-fluid phase.
Here the system is a Luttinger liquid.
The ground state is critical, i.e. the excitation spectrum is
gapless, and the static spin and dimer correlation functions decay
algebraically with exponents depending on the coupling constants of the
two types of interaction.

For $\Lambda <0$ the $nnn$ coupling strengthens the correlations produced by
the $nn$ coupling and thus reinforces the prevailing ordering tendency.
Therefore, we expect to find ferromagnetic long-range order in the region
$(\Lambda \leq 0,\Delta \leq -1)$ and antiferromagnetic long-range order in
the region $(\Lambda \leq 0,\Delta >1)$. For $-1<\Delta \leq 1$ the $nn$
interaction alone is known not to support any kind of long-range order. No
phase transition is suspected to occur if a negative $nnn$ coupling is added.

For $\Lambda >0$ the $nn$ and $nnn$ couplings are in competition with each
other. The latter thus frustrates the ordering tendency of the former.
One predictable consequence is that the boundary of the ferromagnetic phase,
which is located at $\Delta =-1$ for $\Lambda <1/4$, bends to the
left in Fig. \ref{1} for increasing $nnn$ coupling.\cite{BS79}
Increasing amounts of uniaxial anisotropy are necessary to stabilize the
spin alignment.
A simple spin-wave stability criterion yields the expression
$1/\Lambda =4[|\Delta |-\sqrt{\Delta ^2-1}]$ for the phase boundary in that
region.

In the region of planar anisotropy, $|\Delta |<1$, the interplay of the two
competing forces in the presence of quantum fluctuations produces the
dimer phase. Curiously, within the dimer phase there exists one
point in parameter space, $(\Delta =1,\Lambda =1/2)$,\cite{MG69,B80}
where the ground state is a pure dimer state with no correlated
fluctuations, notwithstanding the fact that this phase owes its very
existence to quantum fluctuations.
However, the $T=0$ dynamics at this point turns out to be only marginally
simpler than elsewhere in the dimer phase, where the structure of the ground
state is much more complex.\cite{YM96}

In the region of uniaxial antiferromagnetic anisotropy, $\Delta >1$, there
exists a N\'eel phase. Like the ferromagnetic phase, it is destabilized by
sufficiently strong competing $nnn$ coupling.
The phase boundaries between the spin-fluid, dimer, and N\'eel phases as
sketched in Fig. \ref{1} were first proposed by Haldane\cite{H82} based on
a continuum fermion theory.
The impact of the transitions between the N\'eel, dimer, and spin-fluid
phases on the $T=0$ dynamics will be explored in a separate study.%
\cite{YKM96}

\section{Spin and Dimer Fluctuations}

The N\'eel order parameter (staggered magnetization) and the dimer order
parameter for the model system (\ref{I.1}) are given, respectively,
by the operators
\begin{eqnarray}
\overline{M}_z=\frac 1N&&\sum\limits_{l=1}^N(-1)^lS_l^z\;,
\label{II.1} \\
D=\frac 1N\sum\limits_{l=1}^N(-1)^lD_l\;,\,\,\,&&
D_l=S_l^{+}S_{l+1}^{-}+S_l^{-}S_{l+1}^{+} \;.
\label{II.2}
\end{eqnarray}
N\'eel order in its purest form is realized in the states
\begin{equation}
|\Phi _1^{z}\rangle=|\uparrow \downarrow \uparrow \cdot \cdot
\cdot \downarrow \rangle\;,\,\,\,\,|\Phi_2^{z}\rangle=|\downarrow
\uparrow \downarrow \cdot \cdot \cdot \uparrow \rangle\;,
\label{II.3}
\end{equation}
and pure dimer order in the states
\begin{eqnarray}
|\Phi _1^{D}\rangle&=&[1,2][3,4]...[N-1,N]\;, \nonumber \\
|\Phi_2^{D}\rangle&=&[2,3][4,5]...[N,1]\;.
\label{II.4}
\end{eqnarray}
where the singlets
$[l,l+1]=\{|\uparrow\downarrow\rangle
-|\downarrow\uparrow\rangle\}/\sqrt{2}$
are formed by pairs of $nn$ spins.
In the ground state of (\ref{I.1}), the former is realized at
($\Lambda =0,\,\Delta =\infty$) and the latter at
($\Lambda =1/2,\,\Delta =1$).

N\'eel ordering manifests itself in the two-spin correlation function
$\langle S_l^zS_{l+n}^z\rangle$ and dimer ordering in the four-spin
correlation function $\langle D_lD_{l+n}\rangle$.
For translationally invariant and orthonormal linear combinations of the
symmetry-breaking N\'eel states (\ref{II.3}) and dimer states (\ref{II.4}),
they are $\langle S_l^zS_{l+n}^z\rangle=\frac 14(-1)^n$,
$\langle D_lD_{l+n}\rangle=0$ $(n\neq 0)$, and
$\langle S_l^zS_{l+n}^z\rangle=0$ $(|n|\neq 1)$,
$\langle D_lD_{l+n}\rangle-\langle D_l\rangle\langle D_{l+n}\rangle
= {1 \over 4}(-1)^n$ $(n\neq0)$, respectively, and reflect the long-range
nature of the two types of ordering.
In the free-fermion ground-state at $\Delta =\Lambda =0$, by contrast,
the same correlation functions decay algebraically,\cite{LSM61}
\begin{equation}
\langle S_l^zS_{l+n}^z\rangle=
\frac 1{2\pi ^2n^2}[\cos (n\pi)-1]\;,
\label{II.5}
\end{equation}
\begin{eqnarray}
\langle D_lD_{l+n}\rangle&-&\langle D_l\rangle\langle D_{l+n}\rangle
\nonumber \\
&=&\frac 1{\pi ^2n^2}\left[\frac{2n^2-1}{n^2-1}\cos (n\pi )+
\frac 1{n^2-1}\right]\;.
\label{II.6}
\end{eqnarray}

In this study we use the recursion method\cite{H80,PW85} to investigate the
dynamical (i.e. frequency-dependent)
order-parameter fluctuations as probed by the dynamic spin
structure factor $S_{zz}(q,\omega)$ and the dynamic dimer structure
factor $S_{DD}(q,\omega)$, i.e. by the function
\begin{equation}
S_{AA}(q,\omega)=\int\limits_{-\infty}^{+\infty}dte^{i\omega t}
\langle A_q^\dagger(t)A_q\rangle \;,
\label{II.7}
\end{equation}
where $A_q$ stands for the spin and dimer fluctuation operators,
\begin{eqnarray}
S_q^z&=&N^{-1/2}\sum_le^{iql}S_l^z\;, \nonumber \\
D_q&=&N^{-1/2}\sum_le^{iql}[D_l-\langle D_l\rangle]\;.
\label{II.8}
\end{eqnarray}
For the calculation of the dynamic correlation function
$\langle \Phi|A_q^\dagger(t)A_q|\Phi\rangle$ in the ground
state $|\Phi\rangle$ of the system, it can be based on
an orthogonal expansion of the dynamical variable $A_q(t)$ (Liouvillian
representation)\cite{L82} or on an orthogonal expansion of the wave
function $A_q(-t)|\Phi\rangle$ (Hamiltonian representation).\cite{GB88}
The algorithms of both representations, which have been described and
illustrated in a recent monograph,\cite{VM94} produce equivalent data.
These data are expressible most concisely in terms of a sequence of
continued-fraction coefficients $\Delta_1^A(q)$, $\Delta_2^A(q),\ldots$
for the relaxation function
\begin{equation}
c_0^{AA}(q,z)={\displaystyle {\frac 1{z+{\displaystyle {\frac{\Delta_1^A(q)}
{z+{\displaystyle {\frac{\Delta_2^A(q)}{z+\ldots }}}}}}}}}\;,
\label{II.9}
\end{equation}
which is the Laplace transform of the symmetrized and normalized correlation
function $\Re\langle\Phi|A_q^{\dagger}(t)A_q|\Phi\rangle/$
$\langle\Phi|A_q^{\dagger}A_q|\Phi\rangle$.
The $T=0$ dynamic structure factor (\ref{II.7}) is then obtained from
(\ref{II.9}) via
\begin{equation}
S_{AA}(q,\omega) = 4S_{AA}(q)\Theta(\omega)
\lim \limits_{\epsilon \rightarrow 0}\Re[c_0^{AA}(q,\epsilon-i\omega)]\;,
\label{II.10}
\end{equation}
where $S_{AA}(q)=\langle\Phi|A_q^\dagger A_q|\Phi\rangle$ is the static
structure factor (integrated intensity).

For the reconstruction of $S_{AA}(q,\omega)$ based on a number of
coefficients $\Delta_k^A(q)$ extracted from the finite-size ground-state
wave function $|\Phi\rangle$, we employ techniques of continued-fraction
analysis described previously in the context of other applications -- one
pertaining to the strong-coupling regime\cite{VZSM94} and the other to the
weak-coupling regime\cite{VZMS95} of a given quantum many-body system.

\section{Results and Interpretation}

For the special case $\Lambda =\Delta =0$ of the spin model (\ref{I.1}),
dynamic correlation functions can be calculated exactly.
The evaluation is particularly simple for the two dynamic structure factors
of interest here, because the operators
involved, $S_q^z$ and $D_q$, are density operators in the
fermion representation:\cite{N67,KHS70}
\begin{equation}
S_{zz}(q,\omega )=\frac 2{\sqrt{4J_{\perp}^2\sin^2(q/2)-\omega^2}} \;,
\label{III.1}
\end{equation}
\begin{equation}
S_{DD}(q,\omega)=\frac{\sqrt{4J_{\perp}^2\sin^2(q/2)-\omega^2}}
{J_{\perp}^2\sin^2(q/2)} \;,
\label{III.2}
\end{equation}
for $J_{\perp}|\sin q| < \omega < 2J_{\perp}|\sin (q/2)|$.
Both quantities have the same excitation spectrum $-$ the particle-hole
continuum of free fermions. For other dynamical variables such
as $S_q^x$, the evaluation of dynamic correlation functions is much more
involved, and the results have a much more complicated structure with
dynamically relevant excitation spectra of unbounded support.\cite{MS84}

In the XXZ antiferromagnet with planar anisotropy ($\Lambda =0,\,0\leq
\Delta \leq 1$), the spectral weight in $S_{zz}(q,\omega )$ is
known\cite{MTBB81} to be dominated by a continuum of unbound two-spinon
states with upper and lower boundaries
\begin{equation}
\epsilon_L(q)=J_\Delta |\sin q|\;,\,\,\,\,\,\,
\epsilon_U(q)=2J_\Delta |\sin (q/2)|\;,
\label{III.3}
\end{equation}
where $J_\Delta =\pi J_{\perp }\sin \vartheta /2\vartheta $, cos$\vartheta
=\Delta $.
The $\Delta$-dependence of $S_{zz}(q,\omega)$ has already been investigated
by several calculational techniques.\cite{VZMS95}
No equivalent results exist for $S_{zz}(q,\omega)$ in the extended
parameter space ($\Delta,\Lambda $) or for $S_{DD}(q,\omega)$
anywhere in this parameter space.
Given the exact solutions (\ref{III.1}) and (\ref{III.2}), the case $\Lambda
=\Delta =0$ presents itself as a convenient starting point for a
weak-coupling continued-fraction analysis of the two dynamic structure
factors at $|\Lambda |\ll 1$ and $|\Delta|\ll 1$.\cite{VZMS95}

\subsection{Correlation exponent and spin velocity}

Haldane's continuum analysis\cite{H82} predicts that the exponents which
characterize the power-law decay of the spin and dimer correlation functions
$\langle S_l^zS_{l+n}^z\rangle \sim (-1)^n n^{-\eta_z}$,
$\langle D_lD_{l+n}\rangle-\langle D_l\rangle\langle D_{l+n}\rangle
\sim (-1)^n n^{-\eta_D}$,
are the same: $\eta _z=\eta _D$.
A calculation taking into account both the backscattering and the umklapp
terms in the interaction leads to a pair of scaling equations for
the dependence of $\eta _z$ on $\Delta $ and $\Lambda $.
If the umklapp terms are neglected, the analysis yields the
explicit result:\cite{H82}
\begin{equation}
\eta _z(\Delta ,\Lambda )=2\sqrt{(\pi -8\Lambda )/(\pi+4\Delta )}\;.
\label{III.4}
\end{equation}
This expression is expected to be most accurate for the special coupling
ratio $\Lambda/\Delta=1/6$, where the umklapp terms
are absent in the continuum Hamiltonian.
Indeed the high accuracy of (\ref{III.4}) is demonstrated by the fact that
its value $\eta _z=1.006$ at the boundary to the N\'eel phase
$(\Delta =1,\Lambda =1/6)$ misses the supposedly exact value $\eta _z=1$ by
less than one percent.
It is not clear how accurate expression (\ref{III.4}) is for coupling
ratios $\Lambda /\Delta \neq 1/6$.
At $\Lambda =0$ it can be tested against the exact result\cite{LP75}
\begin{equation}
\eta_z(\Delta,0)_{exact}=[1-(1/\pi)\arccos\Delta]^{-1}\;.
\label{III.5}
\end{equation}
The two expressions agree to leading order in the coupling constant:
$\eta_z=2-4\Delta/\pi+O(\Delta^2)$.

An independent way of determining the correlation exponent $\eta _z$ for
arbitrary coupling ratios $\Lambda /\Delta $ in the region of very weak
interaction is provided, as will be demonstrated in Sec. III.C,
by the weak-coupling continued-fraction analysis of
the infrared exponents in the dynamic spin and dimer structure factors,
\begin{equation}
S_{zz}(\pi ,\omega )\sim \omega ^{\beta_z}\;,\,\,\,\,\,\,\,
S_{DD}(\pi ,\omega )\sim \omega ^{\beta _D}\;.
\label{III.6}
\end{equation}
The continuum analysis suggests that the two infrared exponents are
identical and related to $\eta_z$:
\begin{equation}
\beta_z=\beta_D=\eta_z-2\;.
\label{III.7}
\end{equation}

The renormalized spin velocity $v_s(\Delta,\Lambda)$ is another quantity
for which the continuum analysis predicts an explicit result,\cite{H82}
\begin{equation}
v_s(\Delta,\Lambda ) = J_\perp (1+2\Delta/\pi-4\Lambda/\pi)\;.
\label{III.8}
\end{equation}
It can be checked against the renormalized bandwidth
$\omega_0(\Delta,\Lambda)$ of $S_{zz}(\pi,\omega)$ or $S_{DD}(\pi,\omega)$
as obtained (in Sec. III.C) via weak-coupling continued-fraction analysis.
Both quantities are expected to have the same dependence on the coupling
constants except for constant factors.
At $\Lambda=0$ we have $\omega_0=\epsilon_U(\pi)=2J_\Delta$ and
$v_s=[d\epsilon_U(q)/dq]_{q=0}=J_\Delta$, hence
\begin{equation}
\omega_0^{z} = \omega_0^{D} = 2v_s\;.
\label{III.9}
\end{equation}

\subsection{Weak-coupling continued-fraction coefficients}

The exact expressions (\ref{III.1}) and (\ref{III.2}) can be
recovered for $q=\pi$ from the $\Delta_k$-sequences
\begin{equation}
\Delta_k^z(\pi)=J_{\perp}^2(1+\delta_{k,1})\;,\;\;\;
\Delta_k^D(\pi)=J_{\perp}^2\;,
\label{III.10}
\end{equation}
by direct evaluation of (\ref{II.9}).
These $\Delta_k$ values pertain to the infinite system.
They are exactly reproduced up to $k=N/2-1$ for $S_{zz}(\pi,\omega )$ and up
to $k=N/2-2$ for $S_{DD}(\pi,\omega)$ by the recursion method applied to a
chain of $N$ sites with periodic boundary conditions.

Weak coupling $(|\Delta |\ll 1,\,|\Lambda |\ll 1)$ produces systematic
deviations of the $\Delta_k$'s from the reference sequences (\ref{III.10}).
They are illustrated in the four panels of Fig. \ref{2}
for both dynamic structure factors and both types of interaction. In each
case the $\Delta_k$'s for $K_A\leq k\leq K_W$ exhibit a pseudo-asymptotic
behavior. In panel (a) it starts at $K_A=2$, in panel (b) at $K_A=3$, and in
panels (c), (d) at $K_A=1$. The number $K_W$ marks the beginning of the
crossover from zero growth to power-law growth, which is most conspicuously
observable in panel (d).
$K_W$ becomes smaller with increasing coupling strength. When we have
$K_W\simeq K_A$, we are in the strong-coupling regime.\cite{VZMS95}

In the weak-coupling regime, the systematic deviations from (\ref{III.10})
are of two kinds:
(i) The $\Delta_k$-sequence tends to converge toward a higher
or lower value $\Delta_\infty^{(W)}$ as $k$ increases toward $K_W$.
(ii) The $\Delta_{2k}$ and the $\Delta_{2k-1}$ approach the
pseudo-asymptotic value $\Delta_\infty^{(W)}$ from opposite sides.
In Fig. \ref{2} we observe that the direction of the shift, which changes
with the sign of either interaction, is the same for both dynamic structure
factors but opposite for the two types of coupling, whereas the alternating
pattern, which also changes with the sign of either interaction, has the
same direction for both functions and for both types of coupling.

For a quantitative analysis of the two pseudo-asymptotic effects we must
select the string of $\Delta _k$'s carefully.
In $S_{DD}(\pi,\omega)$, the value $K_W$ is considerably smaller than in
$S_{zz}(\pi,\omega)$ for given coupling strength, which restricts the
weak-coupling analysis of $S_{DD}(\pi,\omega)$
to a narrower range of couplings.

The shift of the pseudo-asymptotic value $\Delta _\infty ^{(W)}$
[effect (i)] describes the renormalized bandwidth $\omega _0$ of the
dynamically predominant 2-particle continuum of lattice fermions via the
relation\cite{VM94}
\begin{equation}
\Delta _\infty ^{(W)}=\omega _0^2/4\;.
\label{III.11}
\end{equation}
For pure $nn$ coupling $(\Lambda =0$, $|\Delta |<0.1)$, the band-edge
frequency $\omega_0$ inferred from the average of $\Delta _2,...,\Delta _5$
was shown to reproduce the exact bandwidth
$\epsilon_U(\pi)=2J_\Delta$ of the spinon continuum (\ref{III.3}) very
accurately.\cite{VZMS95}

The alternating approach of the $\Delta_k$'s toward $\Delta _\infty ^{(W)}$
[effect (ii)] describes an infrared singularity (\ref{III.6}) in the two
dynamic structure factors.
For a truly convergent $\Delta_k$-sequence, the singularity exponent
$\beta$ is governed by the leading term of the large-$k$ asymptotic
expansion:\cite{VM94,M85}
\begin{equation}
\Delta_k=\Delta_\infty [1-(-1)^k\frac \beta k+...]\;.
\label{III.12}
\end{equation}
In Ref. \onlinecite{VZMS95} we proposed and tested two procedures
({\it averaging} and {\it extrapolation}) for extracting the exponent from
a finite $\Delta_k$-sequence.
For the $nn$ case $(\Lambda =0$, $|\Delta |\leq 0.05)$, a benchmark test
showed that the exponent $\beta _z$ inferred from the coefficients
$\Delta _2,...,\Delta _5$ reproduces [via (\ref{III.7})] the exact
correlation function exponent (\ref{III.5}) with reasonable accuracy.

The patterns in Fig. \ref{2} make it clear that in the $(\Delta,\Lambda)$%
-plane there are sectors with compressed $(\Delta \omega _0<0)$ and expanded
$(\Delta \omega _0>0)$ bandwidth, and sectors with divergent $(\beta<0)$
and cusp $(\beta>0)$ singularities.
We have determined these boundaries between the resulting four sectors near
the free-fermion point $(\Delta=0,\,\,\Lambda =0)$ by a
systematic investigation of the $\Delta_k$-sequences along several lines in
parameter space.

One such series of sequences for $S_{zz}(\pi,\omega)$ is displayed in Fig.
\ref{3}.
Between $\Delta =-0.005$ (bottom sequence) and $\Delta=0.006$ (top sequence)
at fixed $\Lambda =-0.001$, we observe a gradual reversal of both patterns
(i) and (ii).
At $\Delta \simeq -0.002$, the average $\Delta_k$ goes from negative to
positive, implying a corresponding change in sign of $\Delta\omega_0^{z}$.
At $\Delta \simeq 0.002$, the average of $\Delta_{2k-1}-\Delta _{2k}$ goes
from positive to negative, implying a corresponding change in sign of the
infrared exponent $\beta _z$.
The two sequences closest to the pattern changes are highlighted by full
symbols.

The lines in parameter space on which the interaction leaves the
bandwidth $\omega_0^{z}$ or the exponent $\beta_z$ unchanged depend
somewhat on the $\Delta_k$-strings used for the analysis.
This is illustrated in Fig. \ref{4}.
The expectation is that the results improve as we shift the
$\Delta_k$-string to higher indices $k$, where non-asymptotic effects become
weaker.\cite{note3}
Our best results, inferred from the string $\Delta_4,\ldots,\Delta_7$,
yield sector boundaries at coupling ratios $\Delta/\Lambda\simeq 0.52$ for
$\Delta\omega_0^{z}=0$ and $\Delta/\Lambda\simeq -0.44$ for $\beta_z=0$,
in fair agreement with the corresponding sector boundaries
$\Delta/\Lambda=0.5$ for $\Delta v_s=0$ and $\Delta/\Lambda=-0.5$ for
$\Delta \eta_z=0$, respectively, predicted by the continuum results
(\ref{III.8}) and (\ref{III.4}) and shown as long-dashed lines in Fig.
\ref{4}.

\subsection{Infrared exponent and bandwidth}

Figure \ref{5}(a) shows the weak-coupling continued-fraction results for
the infrared exponent $\beta_z(\Delta,\Lambda)$ at coupling strengths along
the circle
\begin{equation}
\Delta=\Gamma\cos\theta\;,\;\; \Lambda=\Gamma\sin\theta
\label{III.13}
\end{equation}
with radius $\Gamma=0.01$ around the free-fermion point in the
($\Delta,\Lambda$)-plane.
The three data sets from different $\Delta_k$-strings\cite{note4} fall onto
sine-like curves in this representation, as does the prediction inferred
from the continuum result (\ref{III.4}), shown here as solid line.

The best overall agreement between our data and Haldane's
result occurs at angles $\theta\simeq 9.5^\circ, 189.5^\circ$, which
correspond to the special coupling ratio $\Lambda/\Delta=1/6$ with
vanishing umklapp terms in the continuum Hamiltonian.
However, the agreement between the dashed line and the data set from
$\Delta_4,\ldots,\Delta_7$, which is least affected by non-asymptotic
coefficients, is remarkably good at all angles.

The dependence of the renormalized bandwidth $\omega_0^{z}$ on the same
angular variable $\theta$ is shown in Fig. \ref{5}(b) in a
comparative plot of three sets of weak-coupling continued-fraction data and
the prediction inferred via (\ref{III.9}) from the continuum result
(\ref{III.8}) for the renormalized spin velocity.
Again the agreement improves as the $\Delta_k$-string is shifted toward
higher indices, where it is less affected by non-asymptotic coefficients.

When the exponent and bandwidth are determined for a circle of much smaller
radius, $\Gamma=0.0001$, we find the same angular dependence of
$\beta_z$ and $\omega_0^{z}$ and the same (rescaled) amplitude within
a 5\% margin of error.
This confirms that the $\Delta_k$'s used in this analysis are free of
weak/strong-coupling crossover effects.

We have carried out the same analysis for the renormalized bandwidth
$\omega_0^{D}$ and the infrared exponent $\beta_D$ of $S_{DD}(q,\omega)$.
The results for parameter values on a circular line with radius
$\Gamma=0.0001$ are displayed in Fig. \ref{6}.
This radius had to be chosen much smaller than in the case of
$S_{zz}(\pi,\omega)$ in order to ascertain that at least the
coefficients $\Delta_2,\ldots,\Delta_6$ are free of crossover effects.

The bandwidth data from two different $\Delta_k$ strings depicted in Fig.
\ref{6}(b) are in near perfect agreement with each other and with the result
inferred from (\ref{III.8}).
Likewise the data for the exponent $\beta_D$ as shown in Fig. \ref{6}(a)
exhibits the same characteristic oscillation as already observed in Fig.
\ref{5}(a) for $\beta_z$.
The phase of the oscillation predicted by (\ref{III.4}) is accurately
reproduced by our data.
There are significant deviations in the amplitude,
which are attributable to the strongly non-asymptotic nature of the
coefficients $\Delta_2, \Delta_3$, but the trends indicated by the two data
sets are in the right direction.

The angular dependences of the exponent and bandwidth data shown in
Figs. \ref{5} and \ref{6} are observed to be out of phase by some
$55^\circ$, giving rise to the four sectors of weak-coupling dynamical
behavior discussed in the context of Fig. \ref{4}.
This is consistent with Haldane's predictions (\ref{III.4}) and
(\ref{III.8}) for the correlation exponent and the renormalized Fermi
velocity, respectively.
These quantities have extreme values at $\tan\theta=2$ and $\tan\theta=-2$,
respectively, which corresponds to angles $63.4^\circ$ and
$116.6^\circ=63.4^\circ+53.2^\circ$.

The characteristic oscillations of the infrared-exponent data are also
expected to be present (with the same phase) in the angular dependence of
the static spin and dimer structure factors at the critical wave number
$q=\pi $.
In the free-fermion limit, the Fourier transform of the exact
results (\ref{II.5}) and (\ref{II.6}) yields:
\begin{equation}
S_{zz}(q)=\frac{|q|}{2\pi}\;,\;\;\;S_{DD}(q)=\frac 1\pi[|q|-|\sin q|]\;.
\label{III.14}
\end{equation}
Both functions have a cusp-like maximum at $q=\pi$, which reflects the
critical spin and dimer fluctuations, respectively, in the ground state.
It is reasonable to expect that the variation of
$S_{zz}(\pi)$ and $S_{DD}(\pi)$ with the interaction in the
weak-coupling limit is synchronized with the variation of the correlation
exponent $\eta_z=\eta_D$, and hence with the infrared exponent
$\beta_z=\beta_D$.

In Fig. \ref{7} we have plotted the $\theta$-dependence of $S_{zz}(\pi)$ and
$S_{DD}(\pi)$ for fixed coupling strengths $\Gamma=0.0001$.
Both quantities exhibit the characteristic sinusoidal behavior.
The amplitude is proportional to $\Gamma$ in each quantity.
Interestingly, the phases are different.
Whereas $S_{DD}(\pi)$ varies in phase with the infrared exponent,
$S_{zz}(\pi)$ does not.

\subsection{Reconstruction of $S_{zz}(\pi,\omega)$ and $S_{DD}(\pi,\omega)$.}

The third major feature (in addition to the renormalized bandwidth and the
infrared exponent) characterizing the line shape of the dynamic spin and
dimer structure factors at the critical wave number $q=\pi$ is the
detailed spectral-weight distribution near the band edge.
For the pure $nn$ case ($\Lambda=0$), we found that
$\Delta\neq 0$ causes a redistribution of the spectral weight in
$S_{zz}(\pi,\omega)$ near the band edge and (for $\Delta<0$ only) the
appearance of a discrete spectral line outside the band.\cite{VZMS95}
This discrete state was identified (by Bethe ansatz) to belong to a branch
of bound spin complexes.

For the more general model (\ref{I.1}) with $\Lambda\neq 0$, it is not known
for which parameter values such discrete states exist.
We have investigated this question for parameter values ($\Delta,\Lambda$)
on the circular line (\ref{III.13}) by means of a weak-coupling
continued-fraction reconstruction of $S_{zz}(\pi,\omega)$ based on the
coefficients $\Delta_1,...,\Delta_{K_w}$ and a termination function which
incorporates the renormalized bandwidth $\omega_0^{z}$ and the infrared
exponent $\beta_z$.
A detailed description of the procedure can be found in Refs.
\onlinecite{VM94,VZMS95}.

The dependence on the angular parameter $\theta$ of the
reconstructed $S_{zz}(\pi,\omega)$ is displayed in Fig. \ref{8}.
As $\theta$ increases from zero to $60^{\text{o}}$ [panel (a)], the
bandwidth shrinks and the infrared divergence gains strength.
In addition to these two features we observe the emergence at $\theta\simeq
15^\circ$ of a discrete spectral line above and outside the band.
In the reconstructed relaxation function the discrete state is represented
by an isolated pole.
For $z=\epsilon -i\omega$ with $\epsilon=0.0001$ it has a nonzero width.
At $\theta=15^\circ$, the discrete state is barely distinguishable from
the band edge.

Between $\theta=60^\circ$ and $\theta=120^\circ$ the
bandwidth continues to shrink, while more and more spectral weight is
transferred from the continuum to the discrete state.
The infrared divergence, which has reached its maximum strength at
$\theta\simeq 60^\circ$, weakens over this parameter range.
The further evolution of the line shapes near the band edge and at small
frequencies is shown in panel (b). The infrared divergence disappears at
$\theta\simeq 150^\circ$ and turns into a cusp singularity (see inset).
The bandwidth of the continuum expands while the discrete state moves closer
to the band edge and slowly loses spectral weight.

These trends continue between $\theta=195^\circ$ and
$\theta=240^\circ$ as shown in panel (c).
The depletion of spectral weight at small $\omega$ becomes more and more
prounouced, and the discrete state merges with the band edge at
$\theta\simeq 210^\circ$.
Nothing dramatic happens to the line shape on the last stretch of the
circle. The range of parameter values where a
discrete state is observed, $30^\circ\leq\theta\leq 210^\circ$, roughly
coincides with the range where the renormalized bandwidth is compressed.
In the fermion representation, this is the region, where the Fermi velocity
is renormalized downward.

In the weak-coupling reconstruction of $S_{DD}(\pi,\omega)$,
which starts from the expression (\ref{III.2}) and the
constant $\Delta_k$-sequence (\ref{III.10}), the effects of a weak
interaction (\ref{III.13}) on the line shape are similar in two aspects yet
different in a third aspect.
(i) The low-frequency behavior is governed by a weak infrared singularity
which switches from a divergence to a cusp in accordance with the exponent
data presented in Fig. \ref{6}(a).
(ii) The only noticeable change at the band edge is the variation of the
continuum boundary in accordance with the bandwidth data presented in
Fig. \ref{6}(b).
(iii) However, no noticeable rearrangement of spectral weight near the
band edge occurs unlike what has been observed in $S_{zz}(\pi,\omega)$.
In particular, the discrete state found in $S_{zz}(\pi,\omega)$
does not carry any spectral weight in $S_{DD}(\pi,\omega)$.
It does not play any role in the dimer fluctuations.

\acknowledgments
The work at URI was supported by the U.S. National Science Foundation,
Grant DMR-93-12252, and the work at ORNL by the U.S. Department of Energy
under contract No. DE-FG06-94ER45519. Computations were carried out
at the National Center for Supercomputing Applications,
University of Illinois at Urbana-Champaign.



\begin{figure}
\caption[one]{Zero-temperature phase diagram of the model system (\ref{I.1})
with coupling constants $\Delta=J_z/J_\perp$ and $\Lambda=J_2/J_\perp$
in a schematic representation.}
\label{1}
\end{figure}

\begin{figure}
\caption[two]{Continued-fraction coefficients for the dynamic structure
factors $S_{zz}(q,\omega)$ (left) and $S_{DD}(q,\omega)$ (right) at $T=0$
in the weak-coupling regime with $\Lambda=0$ (top) or $\Delta=0$ (bottom).
The energy unit is $J_\perp=1$.}
\label{2}
\end{figure}

\begin{figure}
\caption[three] {$\Delta_{k}$-sequences for the dynamic structure factor
$S_{zz}(\pi,\omega)$ for $\Lambda=-0.001$, $\Delta =
-0.005, -0.004, \ldots, 0.006$. The energy unit is $J_\perp=1$.}
\label{3}
\end{figure}

\begin{figure}
\caption[four]{The spin-fluid phase in the weak-coupling regime is divided
into four sectors with infrared exponents $\beta_{z}$ and
bandwidth renormalization $\Delta\omega_0^z$ of opposite sign.
The solid lines denote the sector boundaries derived from the strings
$\Delta_2,...,\Delta_5$ (circles), $\Delta_3,...,\Delta_6$ (squares), and
$\Delta_4,...,\Delta_7$ (triangles).
One solid line is extended short-dashed into the weak/strong-coupling
crossover region of coefficient $\Delta_7$.
The long-dashed lines with slope $\Lambda/\Delta=\pm 1/2$ are the sector
boundaries predicted by the continuum analysis.}
\label{4}
\end{figure}

\begin{figure}
\caption[five]{(a) Infrared exponent $\beta_z$ of $S_{zz}(\pi,\omega )$ and
(b) renormalized bandwidth $\omega_0^z$ of the two-spinon continuum at
different coupling ratios $\Lambda/\Delta=\tan\theta$ and fixed coupling
strength $\sqrt{\Delta^2+\Lambda^2}=\Gamma=0.01$.
The squares, circles, and triangles represent the results of a
weak-coupling continued-fraction analysis based on three different
$\Delta_k$-strings.
The solid lines in (a) and (b) are derived from the continuum results
(\ref{III.4}) and (\ref{III.8}), respectively.
The energy unit is $J_\perp=1$.}
\label{5}
\end{figure}

\begin{figure}
\caption[six]{(a) Infrared exponent $\beta_D$ of $S_{DD}(\pi,\omega)$ and
(b) renormalized bandwidth $\omega_0^D$ of the two-spinon continuum at
different coupling ratios $\Lambda/\Delta=\tan\theta$ and fixed coupling
strength $\sqrt{\Delta^2+\Lambda^2}=\Gamma=0.0001$.
The squares and circles represent the results of a
weak-coupling continued-fraction analysis based on two different
$\Delta_k$-strings.
The solid lines in (a) and (b) are derived from the continuum results
(\ref{III.4}) and (\ref{III.8}), respectively.
The energy unit is $J_\perp=1$.}
\label{6}
\end{figure}

\begin{figure}
\caption[seven]{Integrated intensities $S_{zz}(\pi)$ and $S_{DD}(\pi)$ of
the spin and dimer order-parameter fluctuations, respectively at different
coupling ratios $\Lambda/\Delta =\tan\theta$ and fixed coupling strength
$\Gamma=\sqrt{\Delta^2+\Lambda^2}=0.0001$ as derived from the ground-state
wave function for $N=16$.}
\label{7}
\end{figure}

\begin{figure}
\caption[eight]{Spin dynamic structure factor $S_{zz}(\pi,\omega)$ at
different coupling ratios $\Lambda/\Delta=\tan\theta$ and fixed coupling
strength $\sqrt{\Delta^2+\Lambda^2}=\Gamma=0.01$, specifically for the
angles (a) $\theta=0^\circ,15^\circ,\ldots,60^\circ$,
(b) $\theta=135^\circ,150^\circ,\ldots,180^\circ$,
(c) $195^\circ,210^\circ,\ldots,240^\circ$.
The main plot depicts the line shapes near the band edge and the inset the
line shapes at low frequencies.
All curves have resulted from a weak-coupling continued-fraction
reconstruction based on the coefficients $\Delta_1,\ldots,\Delta_7$ and a
compact $\alpha$-terminator as explained in Refs.
\onlinecite{VM94,VZMS95}. The energy unit is $J_\perp=1$.}
\label{8}
\end{figure}


\begin{references}
\bibitem{note3}
Obviously, the weak-coupling continued-fraction results for the infrared
exponent $\beta_z$ and the renormalized bandwidth $\omega_0$ from a string
$\Delta_{k_i},...,\Delta_{k_f}$ of coefficients are vulnerable to
systematic errors.
The two main sources of systematic error are the following:
(i) If $k_i$ is chosen too small ($k_i<K_A$), then the first one or several
$\Delta_k$'s are too strongly non-asymptotic in character and introduce a
bias into the pseudo-asymptotic analysis.
(ii) If $k_f$ is chosen too high ($k_f>K_W$), then the last one or
several $\Delta_k$'s are affected by the crossover to nonzero growth rate,
which introduces a different kind of bias.
\bibitem{note4}
In Ref. \onlinecite{VZMS95} we showed that the accuracy of the
weak-coupling continued-fraction results for the case $\Lambda =0$ can yet
be improved substantially if we use {\it extrapolation} instead of
{\it averaging}.
However, since the data do not lend themselves to extrapolation for all
coupling ratios, we employ the more robust averaging method
for all cases here.
\end{references}
\end{document}